\documentstyle[prb,aps,epsf,twocolumn]{revtex}
\begin{document}
\draft

\title{Mass spectrometric and first principles study of Al$_n$C$^-$ clusters}

\author{Jijun Zhao $^a$$^*$, Bingchen Liu$^{b,c}$, Huajin Zhai$^{b,c}$, Rufang Zhou$^{b,c}$, Guoquan Ni$^{b,c}$, Zhizhan Xu$^b$}

\address{$^a$ Department of Physics and Astronomy, University of North Carolina at Chapel Hill, Chapel Hill, North Carolina 27599-3255 \\ 
$^b$ Laboratory for Quantum Optics, $^c$: Laboratory for High Intensity Optics, Shanghai Institute of Optics and Fine Mechanics, Chinese Academy of Sciences, Shanghai 201800, P.R.China}

\maketitle

\begin{abstract}

We study the carbon-dope aluminum clusters by using time-of-flight mass spectrum experiments and {\em ab initio} calculations. Mass abundance distributions are obtained for anionic aluminum and aluminum-carbon mixed clusters. Besides the well-known magic aluminum clusters such as Al$_{13}^-$ and Al$_{23}^-$, Al$_7$C$^-$ cluster is found to be particularly stable among those Al$_n$C$^-$ clusters. Density functional calculations are performed to determine the ground state structures of Al$_n$C$^-$ clusters. Our results show that the Al$_7$C$^-$ is a magic cluster with extremely high stability, which might serve as building block of the cluster-assembled materials. 

\end{abstract}

\pacs{36.40.Mr, 36.40.Qv, 61.46.+w}

In recent years, clusters and cluster-based materials have been a field of intensive research due to both fundamental and technological importance \cite{1,2,3,4,5,6,7}. The structural, electronic, magnetic, and optical properties of the clusters are different from those of constitute atoms or bulk phase and depend sensitively on the size and composition of the cluster \cite{1,2,3}. It is desirable to assemble the cluster-based materials from properly designed clusters so that the unique properties of these individual clusters can be retained \cite{6,7}. To be a building block of cluster-assembled materials, the cluster should be highly stable and relatively unreactive. Thus, the clusters would interact weakly with each other and maintain their identities when they are brought together in the cluster-assembled solids. A well-known example is the C$_{60}$ solids \cite{8}. Besides C$_{60}$, it has also been suggested that the metal clusters with electronic and geometric shell structure can be used to construct cluster-assembled materials \cite{6}. In this direction, some previous theoretical efforts have been devoted to aluminum-based clusters \cite{6,7,9,10,11,12,13}. First principles calculations predicted that doping in the aluminum clusters can enhance the stability of certain magic clusters like Al$_{13}$ and modify the physical and chemical properties of the clusters \cite{9,11}. The stability and electronic properties of Al$_{13}$K solid has also been calculated \cite{10}. 

Experimentally, particular attention has been paid to pure and carbon-doped aluminum clusters. The mass spectra of pure and doped Al$_n$ clusters have been obtained by different groups \cite{14,15,16,17}. L.S.Wang et al. have performed a combined photoelectron spectroscopy and {\em ab initio} study on the small neutral and anionic aluminum-carbon clusters Al$_n$C$^-$ ($n=3-5$) \cite{18,19,20}. The reaction of Al$_n$C$^-$ clusters with oxygen was examined by Castleman's group and Al$_7$C$^-$ was found as a magic cluster \cite{21}. In this work, we shall investigate the structure and stability of Al$_n$C$^-$ clusters via a combination of experimental cluster mass spectroscopy and {\em ab initio} calculations. 

Our experimental apparatus consists of a standard Smalley-type laser vaporization/molecular beam cluster source, and a Wiley-McLaren time-of-flight mass spectrometer (TOF MS) \cite{22}. We use a Q-switched frequency doubled Nd: YAG laser [532nm, 15ns full width at half maximum (FWHM)] to vaporize the 6 mm graphite target rod. The aluminum target contains a trace amount of carbon with C/Al ratio as 3$\times$10$^{-4}$. The laser spot on the target is less than 1 mm and the typical laser output is about 40 mJ. Pulsed He carrier gas stream (purity: 99.995$\%$) generated from a pulsed valve is used to cool down the plasma above the target that has been produced during the laser vaporization process. The clusters synthesized in the nozzle channel will then be carried into vacuum via supersonic expansion. The high pressure of the He carrier gas is necessary for generating the anionic Al$_n^-$ and Al$_n$C$^-$ clusters in a mass range from $n=3$ through $n=38$. The clusters form a cluster beam after passing a skimmer (2 mm, Beam Dynamics) located about 2 cm downstream from the nozzle exit. 

The cluster mass abundance distribution is analyzed by the Wiley-McLaren TOF MS driven with a negative high voltage pulse generator (EG $\&$ G Model 1211, 100 ns rise time, -2 kV voltage). The extraction and acceleration regions of the TOF MS are 2 cm and 1 cm long respectively. The length of the free-drift tube in the TOF MS is 105 cm. An X-deflector is applied immediately after the acceleration region to compensate the transverse energy of the ionized clusters. By adjusting the voltage exerted on the X-deflector, clusters in different mass ranges can be recorded. The clusters are detected by a dual-microchannel plate (DMCP) and the signal from the DMCP detector is fed into a LeCroy 9350AL digital oscilloscope (500 MHz bandwidth, 1GHz sampling rate and 2$\times$2M maximum record length). Typically, each mass spectrum is obtained by averaging 200 shots or more in our experiments.

In this work, all electron density functional calculations on Al$_n$C$^-$ (n=1-13) cluster anions have been performed by using the DMol program \cite{23}. A double numerical basis including $d$-polarization function (DND) are chosen. The density functional is treated by the generalized gradient approximation (GGA) with exchange-correlation potential parameterized by Wang and Perdew \cite{24}. Self-consistent field calculations are done with a convergence criterion of 10$^{-6}$ a.u. on the total energy and electron density. For each cluster size, we started from a number of possible configurations and perform full geometry optimizations with Broyden-Fletcher-Goldfarb-Shanno (BFGS) algorithm. We use convergence criterion of 10$^{-3}$ a.u on the gradient and displacement, and 10$^{-5}$ a.u. on the total energy in the geometry optimization. For the smaller clusters, different possible spin multiplicities are also tried.

\begin{figure}
\vspace{0.45in}
\centerline{
\epsfxsize=3.0in \epsfbox{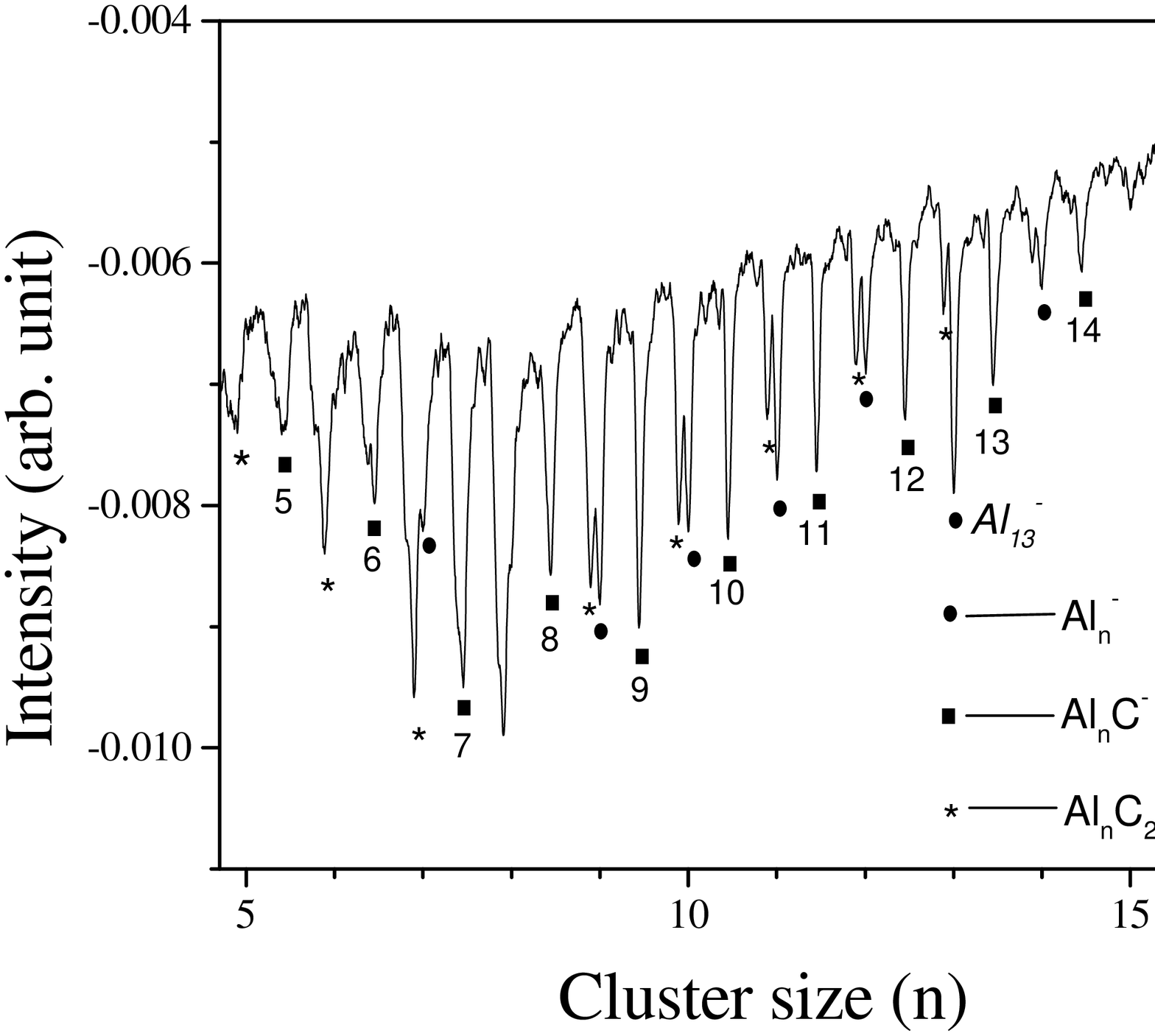}
}
\vspace{-0.75in}
\caption{Mass spectrum for aluminum and aluminum-carbon mixed cluster anions, Al$_n^-$ Al$_n$C$^-$ and Al$_n$C$_2^-$, in the small size range ($n=5-17$). The spectrum peaks for Al$_n$C$^-$ anions and magic Al$_{13}^-$ clusters are marked. Among the Al$_n$C$^-$ clusters, a global maximum at Al$_7$C$^-$ and a less notable local maximum at Al$_9$C$^-$ can be identified. The detailed spectrum structures around Al$_7$C$^-$ are further illustrated in Fig.2.}
\end{figure}

\begin{figure}
\vspace{0.45in}
\centerline{
\epsfxsize=3.0in \epsfbox{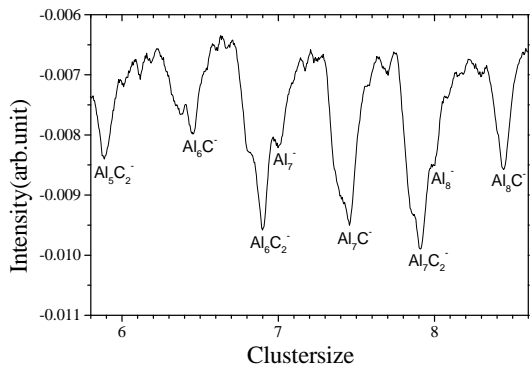}
}
\vspace{-0.75in}
\caption{An expanded spectrum exhibits the more detailed structure of Al$_n^-$, Al$_n$C$^-$ and Al$_n$C$_2^-$ in the mass range from $n=5$ to $n=8$.}
\end{figure}

The spectrum of cluster anions Al$_n^-$ and Al$_n$C$^-$ in Fig.1 covers the small cluster size ranging from $n=5$ to $n=17$. The spectrum is recorded when a helium carrier gas stagnation pressure of 8 atm and a deflection voltage of 90 V exerted on the X-deflector are used. One prominent feature in the spectrum is the well-known magic cluster Al$_{13}^-$, which corresponds to a complete icosahedral geometrical shell structure \cite{9} and a $2p$ electronic shell closing with the total number of valence electrons $n_e=40$ \cite{1,2,16}. In Fig.1, three sequential maximum peaks centered at Al$_7$C$^-$ are also found. For clarity, an expanded spectrum in the mass region around Al$_7$C$^-$ with a better resolution is presented in Fig.2. Al$_6$C$_2^-$, Al$_7$C$^-$ and Al$_7$C$_2^-$ are identified as local maxima. It is noted that under the above experimental conditions, these species are always found as local maxima. Therefore, one can conclude from the mass spectrum that the Al$_7$C$^-$ has the highest abundance among the anionic Al$_n$C$^-$ clusters. The present results on Al$_7$C$^-$ agree well with previous experimental mass spectra \cite{15,21}.

To investigate the aluminum cluster anions in a larger mass range, we use a 10 atm He gas stagnation pressure and a 270 V X-deflection voltage. The spectrum of Al$_n^-$ and Al$_n$C$^-$ anions recorded in such conditions is presented in Fig.3. In Fig.3, Al$_{23}^-$ is obtained as a local maximum. The seventy valence electrons in Al$_{23}^-$ coincide the $3s$ electronic shell closing under the spherical jellium model \cite{1,2,16}. The geometric structure of Al$_{23}^-$ may also contribute to its stability \cite{12,26,27}. Whether the slightly intense Al$_{35}^-$ ($n_e=106$) peak in the spectrum has an electronic shell origin remains an issue of discussion. Besides, a sudden increase in intensity of Al$_{17}^-$ and Al$_{17}$C$^-$ is found in Fig.3. Our present mass spectrum of  Al$_n^-$ compares well with previous results \cite{15}. However, in this study, we would like to pay the particular attention to the unusually stable species Al$_7$C$^-$.

\begin{figure}
\vspace{0.35in}
\centerline{
\epsfxsize=3.0in \epsfbox{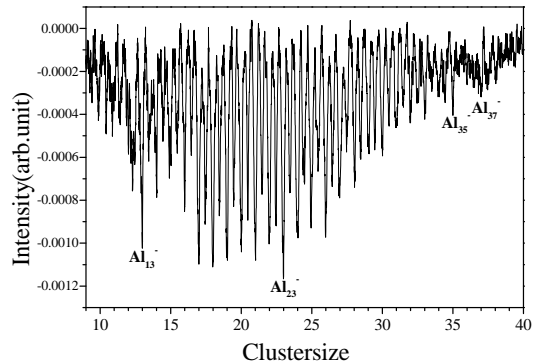}
}
\vspace{-0.75in}
\caption{A mass spectrum of Al$_n^-$ and Al$_n$C$^-$ in the larger size range.}
\end{figure}

In this paper, we exploit the lowest energy structures of the Al$_n$C$^-$ ($n=1-13$) by using density functional geometrical minimization. The obtained ground state structures of Al$_n$C$^-$ ($n=1-7$, $8-13$) are shown in Fig.4 and Fig.5 respectively. In general, the carbon atom is located in the center of the equilibrium structure of Al$_n$C$^-$ clusters. The calculated cluster properties such as atomization energy and HOMO-LUOM gap are described in Table I. Our present results on the cluster energy difference $\Delta E_n$ agree well with previous calculations \cite{13}.

The ground state configurations and bond length parameters of the smaller Al$_n$C$^-$ ($n=1-4$) compare well with previous calculations in Ref.[13] (Fig.4). Spin triple state is found as ground state of AlC$^-$ and its bond length is 1.89 \AA, which agree well with previous calculations \cite{13,28}. The minimum energy structure found for Al$_2$C$^-$ is an isosceles triangle (C$_{\rm 2v}$) with bond length 1.85 \AA$~$ and apex angle $\theta=110.6^{\circ}$, in good agreement with the 1.85 \AA$~$ and 103 $^{\circ}$ obtained from previous {\em ab initio} calculations \cite{13}. The lowest energy structure of Al$_3$C$^-$ is a carbon-centered planar triangular structure with C-Al bond length 1.91 \AA$~$, which is close to that of 1.90 \AA$~$ in previous studies \cite{13,18}. For Al$_4$C$^-$, the lowest energy configuration is obtained as a carbon-centered planar trapeziform. The Al-C distances in Al$_4$C$^-$ are 1.96 and 2.05 \AA$~$, This structure is consistent with that in Ref.[13] but different from the symmetric constraint calculation done by L.S.Wang {\em et al.} \cite{19}. 

\begin{figure}
\centerline{
\epsfxsize=3.5in \epsfbox{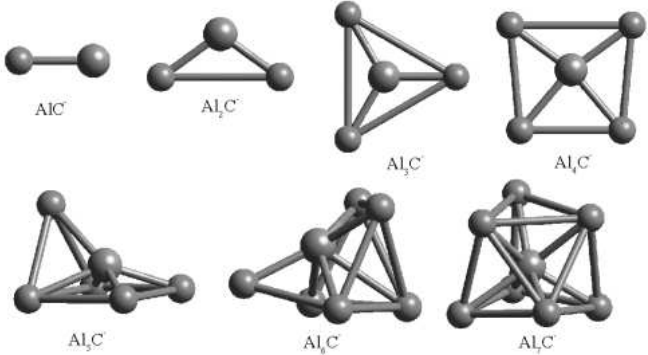}
}
\caption{Lowest-energy structures for small Al$_n$C$^-$ ($n=1-7$) clusters. The atoms with larger radius denote carbon atom.}
\end{figure}

\begin{table}
Table I. Theoretical atomization energy (AE) and HOMO-LUMO Gap ($\Delta$) for anionic Al$_n$C$^-$ clusters with $n=1-13$. $\Delta E_n$ are the differences in atomization energy between cluster with $n$ and $n-1$ atoms for Al$_n$C$^-$ clusters, while $\Delta E_n$$^a$ are those taken from Ref.[13]. 
\begin{center}
\begin{tabular}{ccccc}
  Cluster    &  AE (eV)& $\Delta E_n$ (eV) & $\Delta E_n$$^a$ (eV) & Gap (eV) \\ \hline
 AlC$^-$     &  4.80  &       &      & 0.62  \\
 Al$_2$C$^-$ &  9.20  & 4.40  & 4.46 & 2.40  \\
 Al$_3$C$^-$ &  13.44 & 4.24  & 3.97 & 2.15  \\
 Al$_4$C$^-$ &  15.98 & 2.54  & 2.62 & 1.45  \\
 Al$_5$C$^-$ & 18.60  & 2.62  & 2.38 & 0.80  \\
 Al$_6$C$^-$ & 21.64  & 3.04  & 3.21 & 0.79  \\
 Al$_7$C$^-$ & 25.66  & 4.02  & 3.91 & 1.72  \\
 Al$_8$C$^-$ & 27.52  & 1.86  & 1.94 & 0.36   \\
 Al$_9$C$^-$ & 30.72  & 3.20  &      & 1.48   \\
 Al$_{10}$C$^-$ & 32.95  & 2.23  &    & 0.37   \\
 Al$_{11}$C$^-$ & 35.68  & 2.73  &    & 0.87   \\
 Al$_{12}$C$^-$ & 39.01  & 3.33  &    & 0.21   \\
 Al$_{13}$C$^-$ & 41.46  & 2.45  &    & 0.52   \\
\end{tabular}
\end{center}
\end{table}

The low-energy structure of Al$_5$C$^-$ can be considered as being a distorted square pyramid face-capped by one Al atom. This structure has been obtained as a local minimum in previous study \cite{20}. In our calculation, it is energetically lower than the equilibrium structure (a compressed octahedron) found in Ref.[13] by 0.09 eV. The low-energy structure for Al$_6$C$^-$ can be constructed by face-capping one more aluminum atom on the configuration of Al$_5$C$^-$ in Fig.4. However, a carbon-centered triangular prism was found as ground state structure in Ref.[13]. The energy difference between these two isomers is 0.21 eV from our calculation. In the case of Al$_7$C$^-$, we find a central carbon surrounded by a seven-atom aluminum cage, in complete agreement with previous work \cite{13}. Thus, the particular low reactivity of Al$_7$C$^-$ against oxygen \cite{21} can be partially understood by the close aluminum cage, which can protected the high reactive carbon site from approaching O$_2$ molecules.

\begin{figure}
\centerline{
\epsfxsize=3.5in \epsfbox{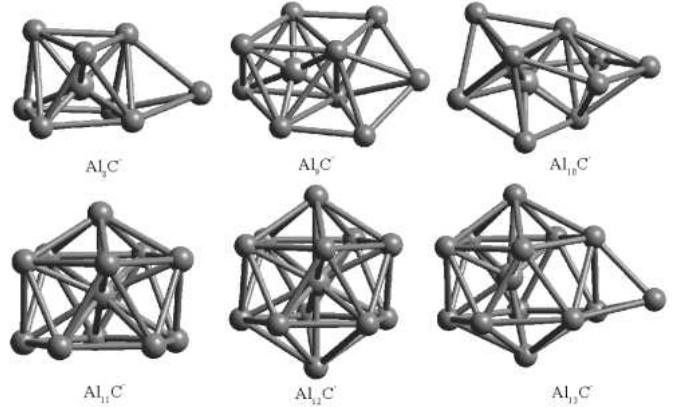}
}
\caption{Lowest-energy structures for larger Al$_n$C$^-$ ($n=8-13$) clusters. The atoms with larger radius denote carbon atom.}
\end{figure}

As shown in Fig.5, the low-energy structures of Al$_8$C$^-$ are based on face-capping of one Al atom on the Al$_7$C$^-$ cage, which agree with previous calculations \cite{13}. The structure for Al$_9$C$^-$ can be viewed as a carbon-centered octahedron edge-capped by three atoms. Similar to Al$_8$C$^-$, the low-energy structures of Al$_{10}$C$^-$ can be constructed by face-capping of three aluminum atoms on the cage-like structure of Al$_7$C$^-$. Starting from $n=11$, the equilibrium configurations of Al$_n$C$^-$ switch to the icosahedron-based packing. Al$_{11}$C$^-$ is a truncated icosahedron with slight distortion. Al$_{12}$C$^-$ is a perfect carbon-centered icosahedron with C-Al distance as 2.56 \AA$~$. The equilibrium structure of Al$_{13}$C$^-$ is a distorted carbon-centered icosahedron capped by one Al atom.

In Fig.6, we plot the second differences of cluster energies defined by $\Delta_2 E (n)=AE(n+1)+AE(n-1)-2AE(n)$, where $AE(n)$ is the atomization energy of Al$_n$C$^-$ clusters from DFT calculations. In cluster physics, the $\Delta_2 E (n)$ is a sensitive quantity that reflect the stability of clusters and can be directly compared to the experimental relative abundance. In Fig.6, a global maximum at $n=7$ and local maxima at $n=3, 9$ are obtained. The present theoretical results are consistent with our experiment and previous works quite well \cite{15,21}. In addition to the most pronounced peak at Al$_7$C$^-$, our experimental mass spectrum in Fig.1 also show a local maxima at $n=9$. On the other hand, Al$_3$C$^-$ was found as a local maximum clusters in Castleman's experiments \cite{21}. 

\begin{figure}
\vspace{0.35in}
\centerline{
\epsfxsize=3.0in \epsfbox{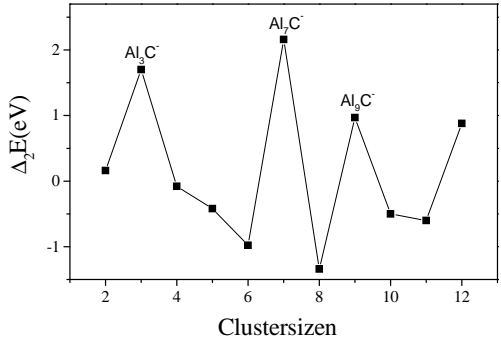}
}
\vspace{-0.75in}
\caption{Second differences of cluster energies $\Delta_2 E(n)$ defined as $\Delta E_2(n)=2AE(n)-AE(n-1)-AE(n+1)$ as a function of cluster size $n$ for $n=2-13$. $E(n)$ is the atomization energy of Al$_n$C$^-$ clusters. Maxima at $n=$3, 7, 9 are found.}
\end{figure}

From our experimental and theoretical investigation, Al$_7$C$^-$ is always obtained as a magic cluster with pronounced high stability among the anionic Al$_n$C$^-$ clusters. In addition to the maximum on the size dependent cluster energy difference, $\Delta E_n$ or $\Delta_2 E(n)$, Al$_7$C$^-$ also possess a HOMO-LUMO gap of 1.72 eV larger than its neighboring clusters (Table I). Such large gap should have certain contribution to the low reactivity upon oxygen. The particular stability of the Al$_7$C$^-$ could be useful in construct cluster-based materials. For example, it might be possible to assemble Al$_7$C$^-$ with alkali metal ions to form stable solid. The further calculations on this direction is still under way and the results will be described elsewhere.

In summary, we have performed both experimental and computational studies on Al$_n$C$^-$ clusters. Experimental mass spectrum on Al$_n^-$ show the standard magic number at Al$_{13}^-$ and Al$_{23}^-$ due to electron shell closing. Among those anionic Al$_n$C$^-$ clusters, Al$_7$C$^-$ is found to be particularly stable. The ground state structures of the Al$_n$C$^-$ ($n=1-13$) clusters are determined from density functional optimizations. Theoretical calculations also demonstrate the high stability of Al$_7$C$^-$, which might be attributed to the stable aluminum cage and large electronic gap. We suggest that the magic Al$_7$C$^-$ cluster could be building blocks of future cluster-assembled materials. 

This work is supported by the U.S. Army Research Office (Grant DAAG55-98-1-0298), NASA Ames Research Center, and the National Natural Science Foundation of China (No.29890210) and National Climbing Project of China. 
 
\ \newline
$^*$ Corresponding author: zhaoj@physics.unc.edu

\end{document}